\documentclass[superscriptaddress,reprint, amsmath,amssymb,aps,prb,longbibliography]{revtex4-2}
\usepackage[utf8]{inputenc}
\usepackage[T1]{fontenc}
\usepackage{dcolumn}
\usepackage{bm}
\usepackage{color,soul}
\usepackage[breaklinks]{hyperref}
\usepackage{listings}
\usepackage{bbold}
\usepackage[dvipsnames]{xcolor}
\usepackage{orcidlink}
\hypersetup{colorlinks=true,linkcolor=blue,citecolor=blue}
\usepackage[nolist,nohyperlinks]{acronym} 

\begin{document}
\title{Tunable  exciton polaritons in biased bilayer graphene}
\author{V. G. M. Duarte\,\orcidlink{0009-0009-5836-6084}}
\email{vgmduarte@gmail.com}
\affiliation{Department of Physics, Aeronautics Institute of Technology, 12228-900, São José dos Campos, SP, Brazil}

\author{P.~Ninhos\,\orcidlink{0000-0002-1143-7457}}
\affiliation{
 POLIMA---Center for Polariton-driven Light--Matter Interactions, University of Southern Denmark, Campusvej 55, DK-5230 Odense M, Denmark
}

\author{C.~Tserkezis\,\orcidlink{0000-0002-2075-9036}}
\affiliation{
 POLIMA---Center for Polariton-driven Light--Matter Interactions, University of Southern Denmark, Campusvej 55, DK-5230 Odense M, Denmark
}

\author{N.~Asger~Mortensen\,\orcidlink{0000-0001-7936-6264}}
\affiliation{
 POLIMA---Center for Polariton-driven Light--Matter Interactions, University of Southern Denmark, Campusvej 55, DK-5230 Odense M, Denmark
}
\affiliation{
 Danish Institute for Advanced Study, University of Southern Denmark, Campusvej 55, DK-5230 Odense M, Denmark
}

\author{N.~M.~R.~Peres\,\orcidlink{0000-0002-7928-8005}}
\affiliation{
 POLIMA---Center for Polariton-driven Light--Matter Interactions, University of Southern Denmark, Campusvej 55, DK-5230 Odense M, Denmark
}
\affiliation{Centro de F\'{\i}sica (CF-UM-UP) and Departamento de F\'{\i}sica, Universidade do Minho, P-4710-057 Braga, Portugal}
\affiliation{International Iberian Nanotechnology Laboratory (INL), Av Mestre Jos\'e Veiga, 4715-330 Braga, Portugal}
\author{A. J. Chaves\,\orcidlink{0000-0003-1381-8568}}
\email{andrejck@ita.br}
\affiliation{Department of Physics, Aeronautics Institute of Technology, 12228-900, São José dos Campos, SP, Brazil}
\affiliation{
 POLIMA---Center for Polariton-driven Light--Matter Interactions, University of Southern Denmark, Campusvej 55, DK-5230 Odense M, Denmark
}

\date{\today}

\begin{abstract}
    By harnessing the unique properties of bilayer graphene, we present a flexible platform for achieving electrically tunable exciton polaritons within a microcavity. Using a semiclassical approach, we solve Maxwell's equations within the cavity, approximating the optical conductivity of bilayer graphene through its excitonic response as described by the Elliott formula. Transitioning to a quantum mechanical framework, we diagonalize the Hamiltonian governing excitons and cavity photons, revealing the resulting polariton dispersions, Hopfield coefficients and Rabi splittings. Our analysis predicts that, under realistic exciton lifetimes, the exciton-photon interaction reaches the strong coupling regime. Furthermore, we explore the integration of an epsilon-near-zero material within the cavity, demonstrating that such a configuration can further enhance the light-matter interaction.
\end{abstract}

\maketitle

\begin{acronym}
\acro{UB}{Upper Branch}
\acro{MB}{Middle Branch}
\acro{LB}{Lower Branch}
\acro{2D}{two-dimensional}
\acro{TE}{transverse electric}
\acro{TM}{transverse magnetic}
\acro{EM}{electromagnetic}
\acro{BSE}{Bethe-Salpeter equation}
\acro{RHS}{right-hand side}
\acro{LHS}{left-hand side}
\end{acronym}

\section{Introduction}

Photonic cavities have been the cornerstones of optical technologies,
ever since lasers were first introduced. Light trapped in such a
cavity---in the simplest case, between two mirrors---enables the
efficient control and modification of the optical properties of a
passive or active element via tailoring of the density of optical
states~\cite{hood_pra64}. This has opened tremendous opportunities
for the realization and growth of the field of cavity electrodynamics,
where the strong coupling of atoms to optical modes allows the
hybridization of the two systems and observation of the corresponding
spectral anticrossing, the so-called Rabi splitting~\cite{thompson_prl68}. The same concept applies equally
well to artificial atoms,
such as quantum dots and wells~\cite{khitrova_natphys2},
while more recently efforts have
been focused on the emission from excitons in
aggregates of organic molecules or
in \ac{2D} semiconductors~\cite{torma_rpp78,tserkezis_rpp83}.
When a semiconductor is placed inside a microcavity,
whose photons have quantized transversal momentum,
the excitons can couple with the cavity photons,
forming the so called exciton polaritons (EPs)~\cite{kavokin2007}.
Polaritons are the result of the interaction of light with any
dipole-carrying material excitation~\cite{basov_nanoph10}, and they
constitute a fertile platform for probing and understanding a great
variety of phenomena, ranging from Bose--Einstein 
condensation~\cite{kasprzak_nature443,amo_nat457} to new topological phases
and light-induced superconductivity~\cite{sentef_sciadv4}. 
The formation of EPs depends on the coupling strength between light and the excitons; in the weak-coupling
regime, the cavity renormalizes the exciton lifetime through
the Purcell effect~\cite{vahala_nature424}, while in the strong- and ultrastrong regimes, the anticrossing of the exciton and cavity modes implies the appearance of two polaritonic branches~\cite{kavokin2007}.

\begin{figure*}[t!]
    \centering
    \includegraphics[width=0.48\columnwidth]{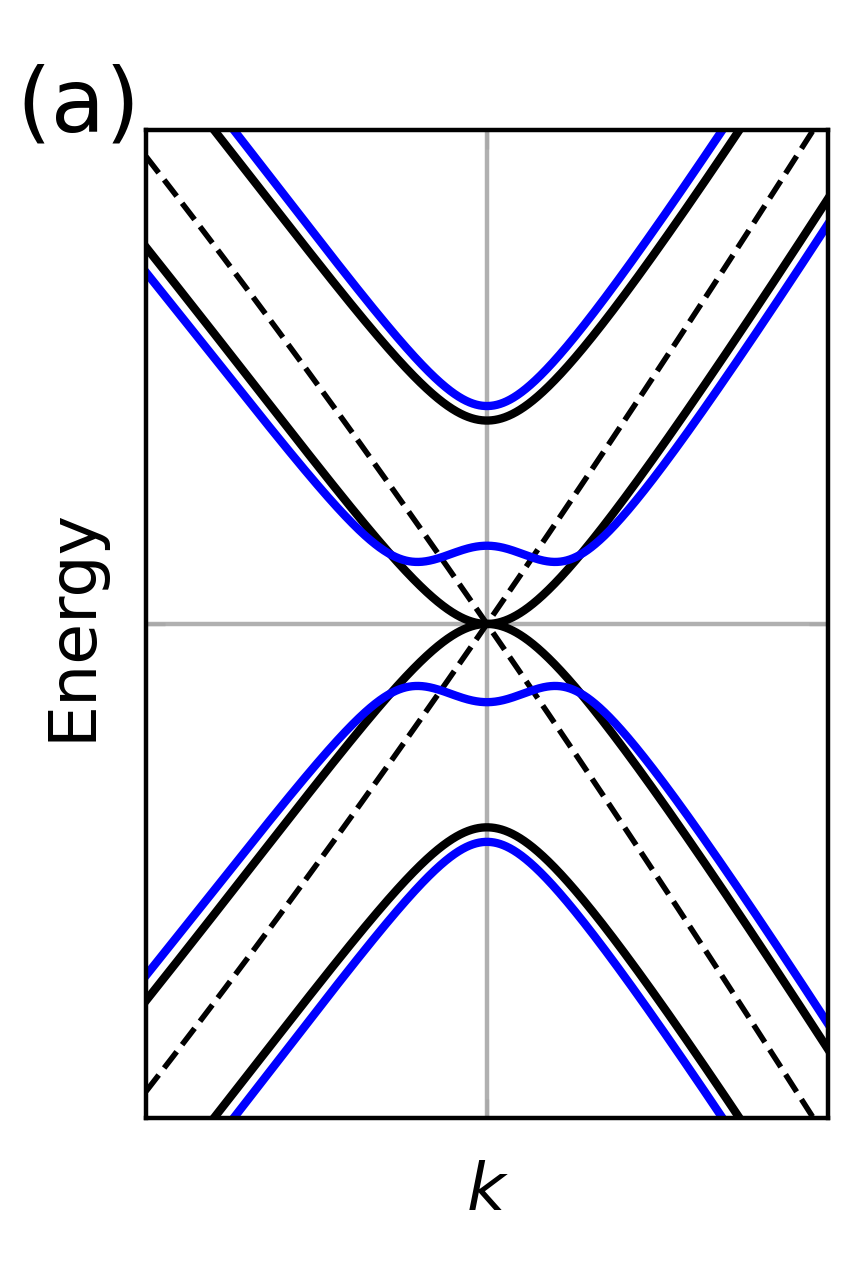}
    \includegraphics[width=0.9\columnwidth]{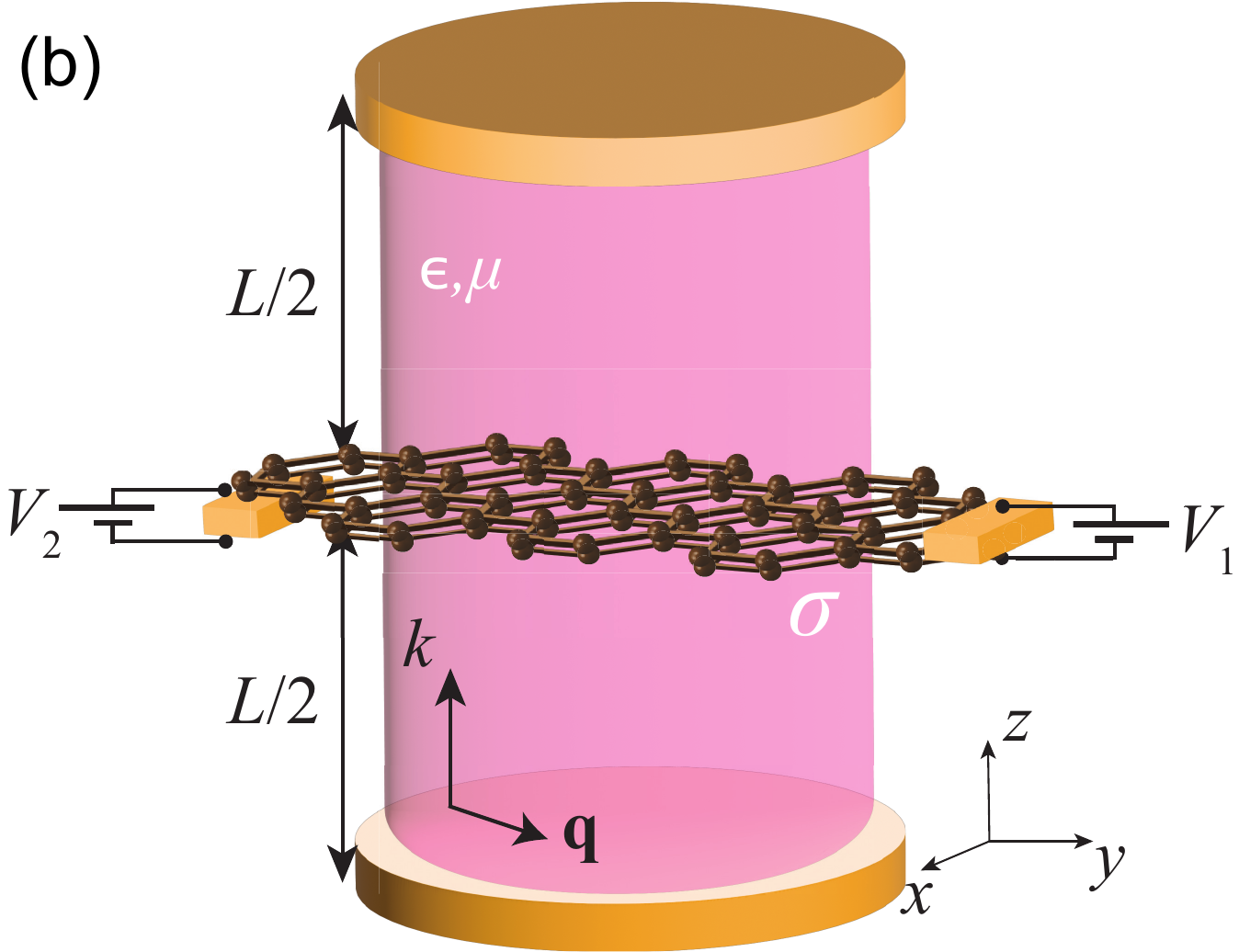}
    \caption{
        (a) Energy dispersion of mono- and bilayer graphene. The blue lines show the gapped dispersion for an applied bias $V_g$ between the graphene sheets, compared to the gapless $V_g=0$ dispersion in solid black lines and the monolayer dispersion in dashed black lines.
        (b) Depiction of a \ac{2D} material (in this case bilayer graphene) 
        with a surface optical conductivity $\sigma(\omega)$ inside a microcavity
        of length $L$, filled with a material of dielectric
        permittivity $\epsilon$ and magnetic permeability $\mu$.
        The cavity plates are assumed to be perfect metals.
        Two electrodes placed on the two ends of the \ac{2D} material ensure an applied interlayer bias $V_{g} = V_{1} - V_{2}$. The wavevector in the \ac{2D} material is decomposed into in-plane ($\mathbf{q}$) and out-of-plane ($k$) components.
        }
     \label{fig1}
\end{figure*}

\ac{2D} materials provide one of the most prominent modern templates for light--matter
interaction~\cite{reserbat-plantey_acsphotonics8}. They owe their popularity to
several reasons, the most significant ones being their rich optical
response over a wide part of the \ac{EM} spectrum, 
the existence of several different polaritonic 
excitations~\cite{basov_sci354}, and
their extreme sensitivity to the dielectric environment. These materials
can be stacked to form different van~der~Waals 
heterostructures~\cite{geim_nature499}, whose physics is drastically impacted
by the interlayer interactions. To give a few examples, monolayer graphene
is a semimetal, whereas bilayer graphene with an electric bias is a
semiconductor~\cite{castro_prl99} (without bias, bilayer graphene is also
a semimetal but, contrary to graphene, has a parabolic dispersion).
In the case of black phosphorus (BP), the opposite occurs: the single
layer material is a semiconductor while biased few-layer BP is a 
metal~\cite{kim_science349}. Recently, the exploration of this stacking in
moir\'{e} semiconductors revealed the presence of localized excitons which
can be electrically controlled, leading to new platforms of tunable quantum-emitter arrays~\cite{yu_sciadv3}. 

The coupling of excitons in \ac{2D} materials to microcavity photons, for transition metal dichalcogenides (TMDs),
was first proposed theoretically by Vasilevskiy \emph{et al.}~\cite{vasilevskiy_prb92},
including the possibility of formation of Bose--Einstein condensates, and experimentally by Schwarz \emph{et al.}~\cite{schwarz_nanolet14}.
Anticrossings of the order of $20$\,meV at room temperature were observed for MoSe$_2$~\cite{dufferwiel_natcom6},
$46$\,meV for MoS$_2$~\cite{liu_naturephotonics9}, and $70$\,meV for WS$_2$~\cite{flatten_scirep6}.
Afterwards, it was also observed for valley-dependent EPs~\cite{chen2017valley},
while Bose--Einstein condensation was evidenced through both luminescence measurements
and via spin-polarizability in an externally applied magnetic field~\cite{Anton-Solanas2021}.
Alternatives to traditional microcavities have been explored in the form
of plasmonic nanostructures~\cite{stuhrenberg_nl18,geisler_acsphot6},
seeking to exploit the near-field enhancement and confinement provided by localized surface plasmons, but at the cost of making the cavity open and more challenging to
characterize~\cite{tserkezis_rpp83}.

In the context of coupling of \ac{2D} materials with photonic cavities, the prototypical \ac{2D} material, namely graphene, has been explored much less.
While graphene and bilayer graphene (without applied bias) are semimetals,
the application of an external vertical electric field can open a 
gap~\cite{castro_prl99,zhang_nature459} in the bilayer case, as depicted in Fig.~\ref{fig1}(a).
Thus, by breaking the symmetry via an electric field, biased bilayer graphene can host excitons~\cite{longju_sci358}.
Microcavity EPs in biased bilayer graphene were conceptually proposed in 2015 by De~Liberato~\cite{deliberato_prb92}, where the
exciton--photon coupling and the possibility of entering the strong and ultrastrong
regimes was discussed. Neglecting the exciton decay rate, it was found that for cavities
with quality factor $Q>10$ it is already possible to resolve polaritonic
resonances. For high-$Q$ cavities, the limiting aspect of the EP will be the exciton non-radiative and dephasing decay rates.
In this work we focus on a cavity with a very
large finesse, implying an associated extremely large quality factor,
approximated as infinite. Within this approximation, we discuss the EP properties based on the extracted exciton lifetimes of bilayer
graphene as calculated in Ref.~\cite{longju_sci358}.

The paper is organized as follows.
In Sec.~\ref{sec:EWM-cavity} we solve the transcendental equation obtained from Maxwell equations whose solution corresponds to the EPs, \emph{i.e.}, a classical picture.
In Sec.~\ref{sec:Ham} we derive the Hamiltonian from the quantization of the undressed \ac{EM} field in the cavity and the interaction with excitons
and compare these result with the classical \ac{EM} approach.
We conclude the paper in Sec.~\ref{sec:conc}. Details of the calculations are given in the Appendices.

\section{Semiclassical formalism: Electromagnetic modes inside a planar cavity} \label{sec:EWM-cavity}

Throughout the paper, we consider a planar optical cavity of length $L$, as depicted in Fig.~\ref{fig1}(b).
A \textcolor{black}{bilayer graphene} is positioned in the middle of the cavity, while an electric bias $V_g$ is enabled between \textcolor{black}{the layers} of the \ac{2D} material~\cite{chakraborty_nl18}.
The cavity is filled with a homogeneous isotropic material of dielectric permittivity $\epsilon=\epsilon_r\epsilon_0$ and magnetic permeability $\mu=\mu_r\mu_0$.
For later use, we also introduce the speed of light in vacuum $c=(\mu_0\epsilon_0)^{-1/2}$ and the impedance $Z=(\mu/\epsilon)^{1/2}=Z_r Z_0$, here expressed in terms of the relative impedance $Z_r=(\mu_r/\epsilon_r)^{1/2}$
and the vacuum impedance $Z_0=(\mu_0/\epsilon_0)^{1/2}$.

\subsection{Transcendental equations}

For the semiclassical approach, we consider cavity photons with \textcolor{black}{EM} fields governed by Maxwell's equations, and boundary conditions imposed by the metallic plates and the \ac{2D} material, with a surface optical conductivity $\sigma(\omega)$.
As detailed in App.~\ref{app:method}, the equation that describes \ac{TE} cavity polaritons is 
\begin{subequations} \label{eq:TE and TM}
\begin{equation}
    \dfrac{k c}{\omega} \cot(kL/2)
    = \dfrac{i\mu_{r}\pi\alpha}{2} \dfrac{\sigma(\omega)}{\sigma_0}, \label{eq:TE}
\end{equation}
while for \ac{TM} modes, we have
\begin{equation}
    \dfrac{\omega}{k c} \cot(kL/2)
    = \dfrac{i\pi\alpha}{2\epsilon_{r}} \dfrac{\sigma(\omega)}{\sigma_0}, \label{eq:TM}
\end{equation}
where $k$ corresponds to the out-of-plane wavenumber of the EP, $\alpha=e^2/(4\pi\epsilon_0\hbar c)$ is the fine-structure constant,
and $\sigma_0=e^2/4\hbar=(\pi/2)\sigma_K$, with $\sigma_K=e^2/h$ being the von~Klitzing constant.
The out-of-plane and in-plane wavenumbers $k$ are $q$ are related to the polariton angular frequency $\omega$ through the dispersion relation of the cavity medium,
\begin{equation}
    \mu\epsilon\omega^2 = k^2 + q^2, \label{eq:pitagoras}
\end{equation}
\end{subequations}
thus making---for symmetry reasons---Eqs.~(\ref{eq:TE}) and (\ref{eq:TM}) identical for a vanishing in-plane momentum.
When $q>0$, however, this polarization degeneracy is broken into \ac{TE} and \ac{TM} modes.

\subsection{Optical conductivity: excitonic contributions} \label{subsec:opt}

For an in-plane isotropic \ac{2D} material, the contributions of the excitons to the optical conductivity can be summarized in the Elliott formula~\cite{Chaves_2017,ElliotOpticalAbsorptionExcitons} [see App.~\ref{app:elliot}]:
\begin{equation}
    \dfrac{\sigma(\omega)}{\sigma_0} = i \sum_{n} \dfrac{p_{n}}{E_{n}} \dfrac{\hbar\omega}{\hbar\omega-E_{n}+i\hbar\gamma_n}, \label{eq:elliot}
\end{equation}
where $\gamma_n$ is a phenomenological constant that modulates the excitonic linewidth.
The exciton lifetime is governed by $1/\gamma_n$.
In the summation, $n$ labels the exciton states, while $E_{n}$ and $p_{n}$ are the corresponding exciton energies and weights, respectively. 
The dimensionless ratio $p_{n}/E_{n}$ measures the oscillatory strength of the exciton.

From now on, we focus on bilayer graphene, a \ac{2D} material whose excitonic physics has been documented both theoretically~\cite{yang_prl103,yang_prb83,park_nanolet10,pengke_prb99} and experimentally~\cite{zhang_nature459,longju_sci358}.
Under vertical bias voltage, Bernal-type (or AB-stacked form) bilayer graphene opens a gap that can be easily controlled,
and the resulting excitons have energies that increase linearly with the bias~\cite{castro_prl99}, as depicted in Fig.~\ref{fig1}(a).

\begin{figure}[t]
\centering
    \includegraphics[width=\columnwidth]{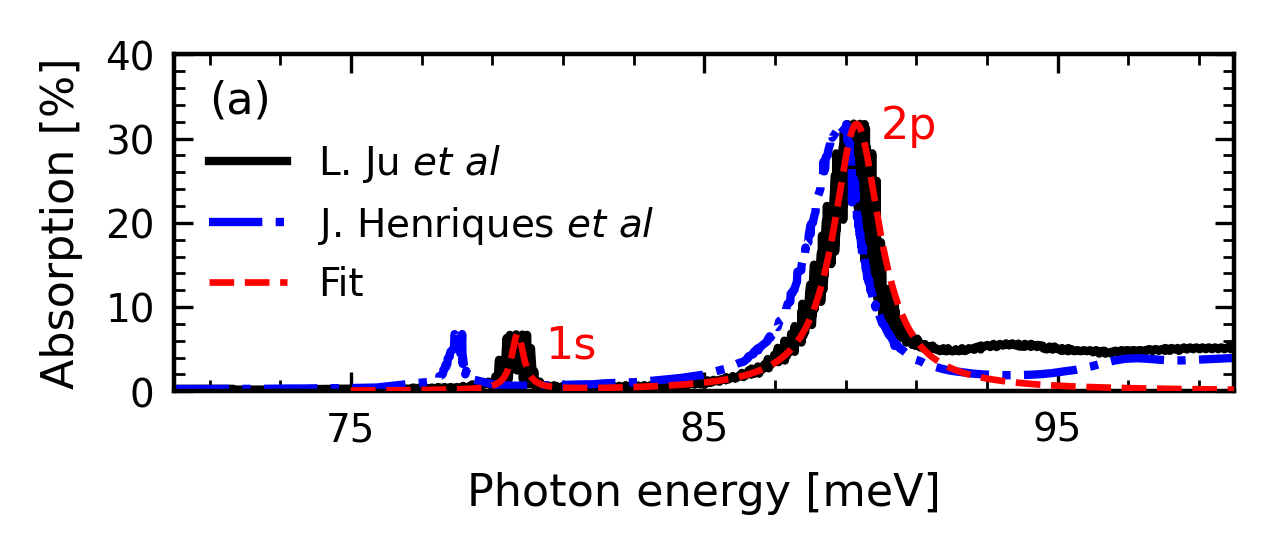}
    \includegraphics[width=\columnwidth]{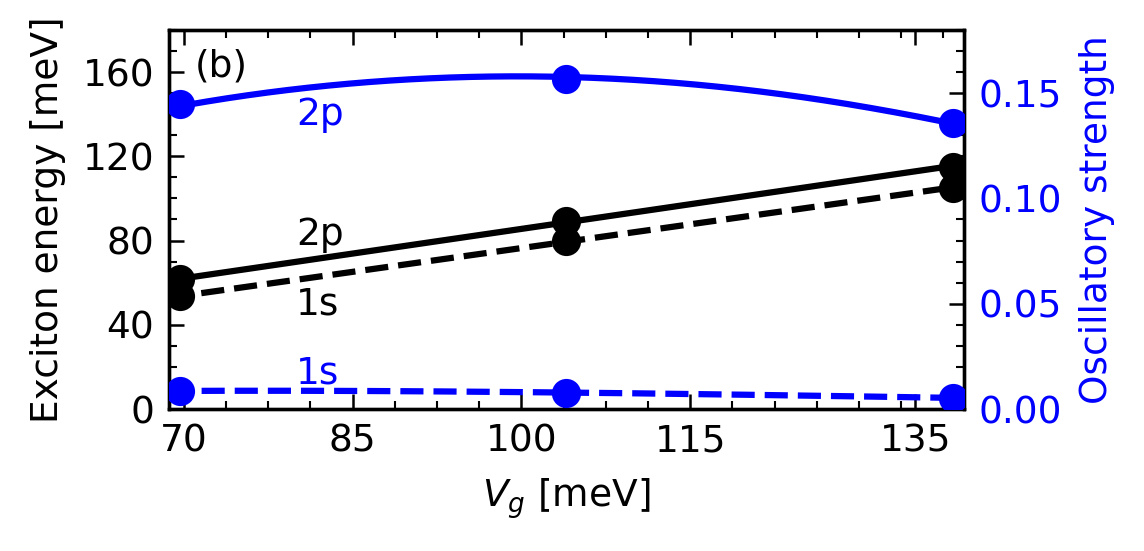}
    \caption{
        Curve fits of the exciton energies and oscillatory strengths with respect to the experimental data from Ref.~\cite{longju_sci358}.
        (a) Absorption spectrum of bilayer graphene, as a function of the incident photon energy.
        The peaks indicate resonance with the $1s$ and $2p$ excitons, with the higher peak corresponding to the $2p$ exciton, indicating stronger coupling.
        Black: experimental data from Ref.~\cite{longju_sci358} for a displacement field of $D=1.03$ V/nm. Red: fits performed as described in Sec.~\ref{subsec:opt}, for $V_g = 104$ meV. Blue: absorption data from Ref.~\cite{nuno_prb105}, obtained through \ac{BSE} calculations.
        (b) Exciton energy $E_n$ (left vertical axis) for bilayer graphene,
        and the oscillatory strength $p_{n}/E_{n}$ (right vertical axis), as function of the bias $V_g$, obtained from the fits.
        The solid (dashed) lines correspond to exciton $2p$ ($1s$) fits.}
    \label{fig2}
\end{figure}

In Fig.~\ref{fig2}(a), we show the experimental photocurrent, which is proportional to the optical absorption, of Ref.~\cite{longju_sci358} in black, 
and in red the theoretical modeling using the Elliott formula~(\ref{eq:elliot}) as explained in the following.
For all the plots in Fig.~\ref{fig2}, we use the same experimental linewidths $2\hbar\gamma_1=0.40$\,meV and $2\hbar\gamma_2=1.3$\,meV.
Ref.~\cite{longju_sci358} uses the displacement field $D$ as a parameter, and a sample of bilayer graphene sandwiched between dielectrics whose widths were not informed.
Since additional absorption data on this system is lacking in the literature, it is not possible to establish a relation between $D$ and the bias $V_g$ without recurring to fits or similar procedures.
Following the approach of Ref.~\cite{nuno_prb105}, which also uses the observations of Ref.~\cite{longju_sci358} to match its theoretical model with experiments, in Fig.~\ref{fig2} we chose the bias $V_g=104$ meV, which best describes the experimental results corresponding to a displacement field $D=1.03$ V/nm.
This correspondence defines a proportionality ratio between $V_g$ and $D$, which allowed us to evaluate the biases that best describe the additional experimental data for $D=0.69$ V/nm ($V_g=70$ meV) and $D=1.37$ V/nm ($V_g=138$ meV).
The absorption fits for this two bias values also exhibited excellent agreement with the correspondent experimental results, but they are omitted since the resultant graphs were very similar to the case of $V_g=104$ meV.
It is worth noting, nonetheless, that this approach assumes that $V_g$ and $D$ are proportional quantities.
This implies that, at $D = 0$\,V/nm, we should have $V_g=0$\,meV and the bandgap should close completely.
In an experimental setting, however, there is an initial offset voltage $V_g^0$ between the graphene sheets~\cite{zhang_nature459}.
However, since the experimental data from Ref.~\cite{longju_sci358} was obtained using an ultraclean sample, this effect was minimized to only a few meV and is therefore negligible for the purposes of this paper.

Unlike traditional semiconductors,
the $2p$ exciton state in bilayer graphene presents the largest oscillatory strength,
an optical transition allowed due to a pseudospin winding number of 2~\cite{longju_sci358}.
The $1s$ exciton state, which is lower in energy, is not dark due to the breaking of rotational symmetry caused by trigonal warping at the band edges.
However, it has a significantly lower oscillatory strength.
In Fig.~\ref{fig2}(b), we show the fit of the experimental values for the exciton energies
and oscillatory strengths, which shows that while the exciton energy increases linearly with the bias, the oscillatory strength behavior is not trivial.
The exciton energies are obtained in a straightforward manner by performing a linear fit $E_n(V_g) = a_1 V_g + a_0$.
For the oscillatory strength $p_n$, conversely, we employed a cubic spline interpolation $\mathcal{A}_{\text{peak},n}(V_g)$ for the absorption peaks $\mathcal{A}_{\text{peak},n}$ ($n=1s,2p$), and used Eq.~(\ref{eq:fit peso}) to obtain the oscillatory strength profile with respect to $V_g$:
\begin{equation}\label{eq:fit peso}
    p_n(V_g) = \dfrac{2\hbar\gamma_n}{\alpha\pi} \dfrac{1-\mathcal{A}_{\text{peak},n}-\sqrt{1-2\mathcal{A}_{\text{peak},n}}}{\mathcal{A}_{\text{peak},n}}.
\end{equation}
This relation between oscillatory strength and absorption is derived from the approximation $\sigma(E_{n}/\hbar)\approx \sigma_0 p_n / (\hbar\gamma_{n})$, which is valid for sufficiently separated excitons in energy, inserted in the absorption expression: 
\begin{subequations}
\begin{equation}\label{eq:absorption}
    \mathcal{A}(\omega) = \dfrac{4\alpha\pi \Re \{\sigma(\omega)\}/\sigma_{0}}{|2 + \alpha\pi \sigma(\omega)/\sigma_0|^2},
\end{equation}
which was calculated using $\mathcal{A}(\omega) = 1 - |\mathcal{R}(\omega)|^2 - |\mathcal{T}(\omega)|^2$ and the reflection and transmission coefficients~\cite{grapheneplasmonics}:
    \begin{eqnarray}
        \mathcal{R}(\omega) = \dfrac{\alpha\pi\sigma(\omega)/\sigma_0}{2+\alpha\pi\sigma(\omega)/\sigma_0},
\\
        \mathcal{T}(\omega) = \dfrac{2}{2+\alpha\pi\sigma(\omega)/\sigma_0}.
    \end{eqnarray}
\end{subequations}

\subsection {Cavity exciton polaritons}

The solution of Eqs.~(\ref{eq:TE}, \ref{eq:TM}) requires careful consideration of the ranges of interest for the frequency $\omega$ and in-plane wavenumber $q$.
We start by rewriting the optical conductivity as
\begin{equation}
    \frac{\sigma(\omega)}{\sigma_0} = \omega\sum_n \frac{p_n}{E_n} \frac{\gamma_n + i(\omega-\textcolor{black}{\omega_n})}{\gamma_n^2 +(\omega-\textcolor{black}{\omega_n})^2},
\end{equation}
\textcolor{black}{where $\omega_n = E_n/\hbar$ are the exciton frequencies. This rearrangement} makes it clear that the imaginary part dominates, and the real part only has relevant influence for frequencies $\omega$ very close to \textcolor{black}{$\omega_n$}.
Therefore, it is reasonable to start by neglecting this contribution.
In this regime, one can notice that the \ac{RHS} of both Eqs.~(\ref{eq:TE}, \ref{eq:TM}) is the same and goes to $0$ at $\omega\rightarrow 0$
and $\sum_{n} p_{n}/E_{n} \approx 0$ at $|\omega-\textcolor{black}{\omega_n}| \gg \gamma_{n}$.
At these limiting cases, we must have $\cot(kL/2)=0$ for the \ac{LHS}, thus $k=n\pi/L$ (with $n=1, 3, 5,\ldots$),
retrieving back the odd-symmetry empty cavity modes [see Eq.~\eqref{eq:empty cavity photon modes}].

\begin{figure}[!ht]
    \includegraphics[width=0.49\textwidth]{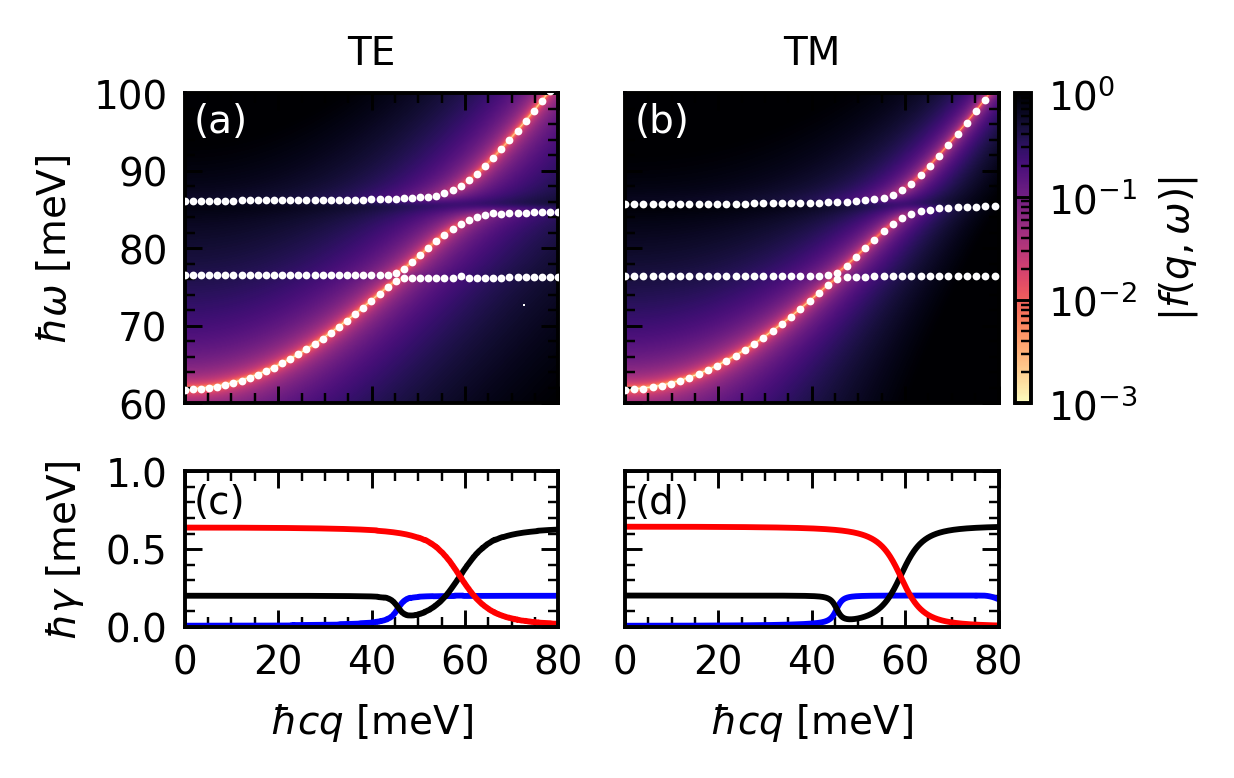}
        \caption{ EP dispersion relations for bilayer graphene in a microcavity of width $L=10\,\mu$m and bias $V_g=100$\,meV filled with air ($\epsilon_r=\mu_r=1$). Color plots of $|f(q,\omega)|$, where $f(q,\omega)$ is the function whose roots correspond to the solutions of Eqs.~(\ref{eq:TE},\ref{eq:TM}), for (a,c) \ac{TE} and (b,d) \ac{TM} modes. Brighter regions indicate where $|f(q,\omega)|$ gets closer to zero. Solutions of Eqs.~(\ref{eq:TE},\ref{eq:TM}) for real $q$ and complex frequency $\omega-i\gamma$, with (a,b) $\omega$ presented in white dots and (c,d) $\gamma$ presented in blue, black and red lines, corresponding to the lower, middle and upper polariton branch in (a,b) panels, respectively.}
        
    \label{fig3}
\end{figure}

Close to the exciton frequencies \textcolor{black}{$\omega_{n}$}, however, the real part of the conductivity becomes relevant, and the dispersion must deviate away from the roots of the cotangent.
Fig.~\ref{fig3} shows a color plot of $|f(q,\omega)|$, where $f(q,\omega)$ is the function whose roots correspond to Eq.~\eqref{eq:TE} or~\eqref{eq:TM}:%
\begin{subequations} \label{eq:fTE and fTM}
    \begin{align}
        f^{\text{TE}}(q,\omega) &= \dfrac{k c}{\omega} \cot(kL/2) - \dfrac{i\mu_{r}\pi\alpha}{2} \dfrac{\sigma(\omega)}{\sigma_0}, \label{eq:fTE}
        \\
        f^{\text{TM}}(q,\omega) &= \dfrac{\omega}{k c} \cot(kL/2) - \dfrac{i\pi\alpha}{2\epsilon_{r}} \dfrac{\sigma(\omega)}{\sigma_0}, \label{eq:fTM}
    \end{align}
\end{subequations}
for real $q$ and $\omega$ as well as a single choice of parameters $L$, $\mu$, $\epsilon$, and $V$.
The brighter regions (\textit{i.e.}, where $f(q,\omega)$ gets close to zero) indicate the best candidate solutions for Eqs.~(\ref{eq:TE}, \ref{eq:TM}).
The numerical minimization of the target function $|f(q,\omega)|$ \textcolor{black}{for complex frequency $\omega\rightarrow \omega-i\gamma$, parameterized by $q$}, is also presented in Fig.~\ref{fig3} \textcolor{black}{[see App.~\ref{app:numerical minimization} for details]}.
This physically corresponds to uniformly illuminating the entire sample for a finite duration.
We denote the solutions of this minimization process as \textcolor{black}{$\{\omega_{i}^{j}(q),\gamma_{i}^{j}(q)\}$, for each range of frequencies $\omega < \omega_{1s}$ ($i=1$), $\omega_{1s} < \omega < \omega_{2p}$ ($i=2$), and $\omega > \omega_{2p}$ ($i=3$).}
As expected, the solutions approach the $n=1$ empty cavity mode for $\omega$ far from \textcolor{black}{$\omega_{1s}$} and \textcolor{black}{$\omega_{2p}$}.
Close to the exciton energies, however, the solution \textcolor{black}{curves $\{\omega_{i}^{j}(q),\gamma_{i}^{j}(q)\}$} deviate from the empty cavity modes around the resonant frequencies \textcolor{black}{$\omega_{n}$}, forming pairs of mutually avoidant curves whose separation increases with $\gamma_{n}$ and $p_{n}$.

Figs.~\ref{fig3}(c) and~\ref{fig3}(d) present the imaginary part of the polariton frequency, which can be interpreted as the decay rate of the EP (with opposite sign).
One can see two crossings at the $q$ values where the polariton dispersion is close to resonance with the excitons.
This shows that the EP decay rate assumes values strictly between the empty cavity polariton decay rate (zero by assumption) and the excitons decay rate ($\hbar\gamma_{1s}=0.20$\,meV and $\hbar\gamma_{2p}=0.65$\,meV),  and assumes half the value of the exciton decay rate at the exciton energy.

\section{Quantum formalism: Hopfield Hamiltonian } \label{sec:Ham}

To investigate the EP modes further, we introduce the Hopfield Hamiltonian (see App.~\ref{app.electron-cavity photon})
\begin{eqnarray}
    H = \sum_{\mathbf{q} n} \hbar\omega_{qn} b^{\dagger}_{\mathbf{q}n} b_{\mathbf{q}n} + \sum_{\mathbf{q}m} E_{m} c^{\dagger}_{\mathbf{q}m} c_{\mathbf{q}m} \nonumber\\
    + \sum_{\mathbf{q}mn} g_{mn}(q) c^{\dagger}_{\mathbf{q}m} b_{\mathbf{q}n} + g^*_{mn}(q) b^\dagger_{\mathbf q n} c_{\mathbf q m}, \label{eq:Hopfield Hamiltonian}
\end{eqnarray}
where $b_{\mathbf{q}n}$ and $c_{\mathbf{q}m}$ are the annihilation operators for the cavity photons and excitons, respectively,
and
\begin{equation}
    \omega_{qn}=\frac{1}{\sqrt{\mu\epsilon}}\sqrt{q^2 + \left(\frac{n\pi}{L}\right)^2}, \label{eq:omegaqn}
\end{equation}
is the empty cavity photon frequency.
The interaction between a cavity photon mode $n$ and an exciton mode $m$ is described by the coupling function $g_{nm}(q)$, whose detailed derivation is presented in App.~\ref{app:elliot}.
To ensure transparency, we adjust the cavity width $L$ so that only the fundamental photon mode $n=1$ resonates with the excitons, and drop the $n$ index from now on for simplicity.
If the excitons are sufficiently separated in energy, we can treat the EPs as decoupled and project the Hopfield Hamiltonian on the subspace of each EP state $|b_{\mathbf{q}}\rangle \otimes |c_{\mathbf{q}m}\rangle$ ($m=1s,2p$), resulting in
\begin{eqnarray}
    H^{\text{eff}}_{m} = \sum_{\mathbf{q}} \bigg\{\hbar \omega_{q} b^{\dagger}_{\mathbf{q}} b_{\mathbf{q}} + E_{m} c^{\dagger}_{\mathbf{q}m} c_{\mathbf{q}m}
    \\
    + g_{m}(q) c^{\dagger}_{\mathbf{q}m} b_{\mathbf{q}} + g_{m}^{*}(q) b^{\dagger}_{\mathbf{q}} c_{\mathbf{q}m} \bigg\}. \label{eq:effective Hopfield Hamiltonian}
\end{eqnarray}
This separation of EPs will create two pairs of curves for each polariton, \emph{i.e.}, one \ac{LB} and one \ac{UB}.
As will become clear further on, the \ac{MB} of the combined system [see Fig.~\ref{fig3}b]
is effectively a mixture of the \ac{UB} of the $1s$ EP and the \ac{LB} of the $2p$ EP.
As discussed in Apps.~\ref{app.electron-cavity photon} and \ref{app:elliot}, the coupling function is given by
\begin{subequations}
    \begin{eqnarray}
        g_{m}^{\text{TE}}(q) = -\hbar \sqrt{\omega_c\omega_{q}} \sqrt{\alpha Z_r \frac{p_m}{E_m}}, \label{eq:gTE}
        \\
        g_{m}^{\text{TM}}(q) = i\hbar \sqrt{\frac{\omega_c^3}{\omega_{q}}} \sqrt{\alpha Z_r \frac{p_m}{E_m}}, \label{eq:gTM}
    \end{eqnarray}
\end{subequations}
where $\omega_c=\pi/(L\sqrt{\mu\epsilon})$ is the fundamental polariton (and empty cavity photon) frequency.

\begin{figure}[ht]
    \includegraphics[width=\columnwidth]{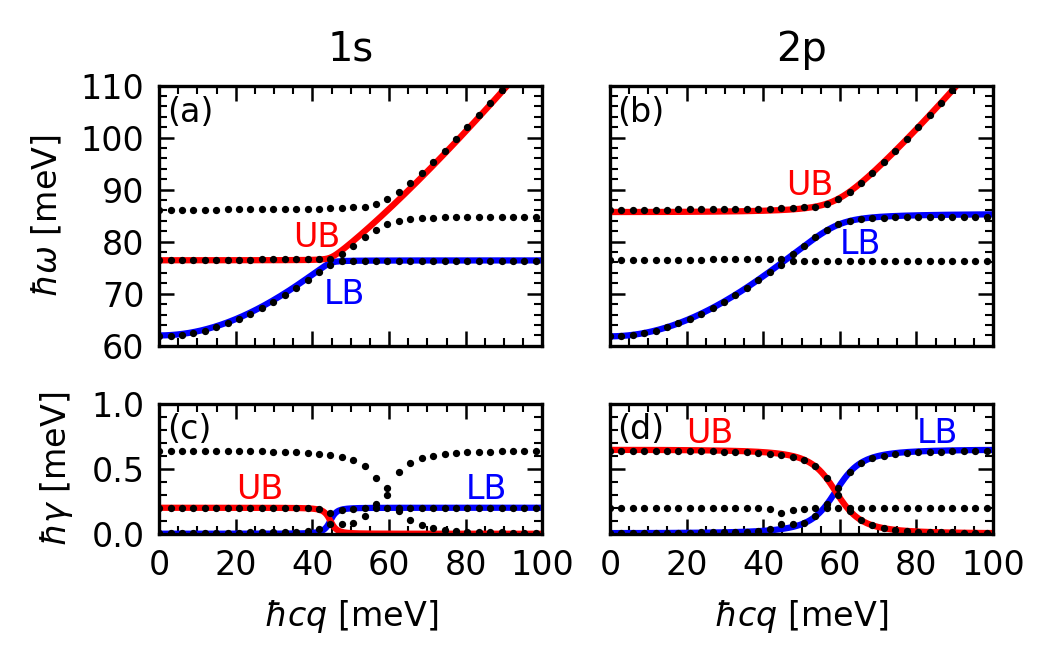}
    \caption{
        EP dispersion relation of \ac{TE} modes for bilayer graphene in a microcavity of width $L=10\,\mu$m
        and bias $V_g=100$\,meV filled with air ($\mu_r=\epsilon_r=1$). All panels show the results of the quantum formalism (lines), in comparison with the classical formalism (dots) as detailed in Fig.~\ref{fig3}.
        (a,b) Eigenenergies of the Hopfield Hamiltonian [see Eq.~\eqref{eq:effective Hopfield Hamiltonian}], for the excitons (a) $1s$ and (b) $2p$.
        (c,d) Effective polaritonic linewidths [see Eq.~\eqref{eq:effective gamma}], for the excitons (c) $1s$ and (d) $2p$.}
    \label{fig4}
\end{figure}

Defining the polaritonic operator
\begin{equation}
    \alpha_{\mathbf{q}m}^j = X^{j}_{qm} c_{\mathbf{q}m} + Y^{j}_{qm} b_{\mathbf{q}}, \label{eq:polaritonic basis}
\end{equation}
and imposing that the effective Hopfield Hamiltonian~\eqref{eq:effective Hopfield Hamiltonian} should be diagonal on this basis, we get
\begin{equation}
    H_{m}^{\text{eff}} = \sum_{\mathbf q m j} \varepsilon^j_{m}(q) \alpha^{j\dagger}_{\mathbf q m} \alpha_{\mathbf q m}^j,
\end{equation}
where $X^j_{qm}, Y^j_{qm}$ are the Hopfield coefficients.
In this setting, the polariton eigenenergies $\varepsilon_{m}^j(q)$ and eigenstates $\{X^j_{q m}, Y^j_{q m}\}$ are the solutions of the secular equation
\begin{equation}
    \begin{pmatrix}
        E_m - \varepsilon_{m}^j(q) & g_{m}(q)\\
        g_{m}^*(q) & \hbar \omega_{q} - \varepsilon^j_{m}(q)
    \end{pmatrix}
    \begin{pmatrix}
        X_{q m}^j \\
        Y_{q m}^j
    \end{pmatrix}
    = 0, \label{eq:secular equation}
\end{equation}
which yields two polariton branches for each EP, labelled by $j$, also denominated the \ac{UB} and \ac{LB}.
Solving the secular equation~\eqref{eq:secular equation}, we get
\begin{equation}\label{eq:Hopfield eigenenergies}
    \varepsilon_{m}^{\pm}(q) = \frac{E_m + \hbar\omega_q \pm \Delta_m(q)}{2},
\end{equation}
where $\Delta_m(q) = [(E_m-\hbar\omega_q)^2 + 4|g_m(q)|^2]^{1/2}$.
The Hopfield coefficients become
\begin{subequations}
    \begin{align}
        X^{\pm}_{q m} &\propto g_m(q),
        \\
        Y^{\pm}_{q m} &\propto \frac{-(E_m-\hbar\omega_q) \pm \Delta_m(q)}{2},
    \end{align}
\end{subequations}
with prefactors defined by the normalization condition $|X^j_{q m}|^2 + |Y^j_{q m}|^2 = 1$.
The eigenenergies~\eqref{eq:Hopfield eigenenergies} profile with respect to $q$ form the dispersion relations of the EP modes,
which are presented in Fig.~\ref{fig4} in comparison with the solution pairs $\{q_0,\omega_0-i\gamma_0\}$
of the transcendental equations from the classical formalism.
As anticipated by the results from the last Section, the \ac{TE} and \ac{TM} mode results were qualitatively analogous, so only the \ac{TE} dispersion of each exciton is presented.
The \ac{MB} is, as expected, a mixture of the \ac{UB} and \ac{LB} of the decoupled $1s$ and $2p$ EPs, respectively, which can be seen by the agreement between the curves of Fig.~\ref{fig4}.
This result shows that the $1s$ and $2p$ EPs are far-apart enough in energy to be treated as effectively decoupled,
and validates the Hopfield model presented in this Section as a good description of the optically dominant polaritons in biased bilayer graphene.

Moreover, the square module of the Hopfield coefficients yield the contribution of the exciton and photon to the polaritonic mode.
Using the Hopfield coefficients and Fermi's golden rule, we obtain
\begin{equation}
    \gamma^{j}_{m}(q)=|X^j_{q m}|^2 \gamma_{m}, \label{eq:effective gamma}
\end{equation}
which establishes a relationship between the exciton linewidths and the so-called effective linewidth of the EP.
Clearly, Fig.~\ref{fig4} shows excellent agreement between $\gamma_{m}^{j}(\mathbf q)$ and the imaginary part of the dispersion obtained previously through Eqs.~(\ref{eq:TE}, \ref{eq:TM}).
This further corroborates the equivalence between the treatment of EPs in bilayer graphene through Eqs.~(\ref{eq:TE}, \ref{eq:TM}) and the Hopfield Hamiltonian for decoupled EPs~\eqref{eq:effective Hopfield Hamiltonian}.
The polariton dispersions form pairs of curves that deviate from the isolated behavior of the exciton and the photon.
Near the region where the dispersions of the isolated exciton and photon would intersect, they instead form pairs of mutually avoidant continuous curves,
with a separation modulated, in essence, by the oscillatory strength of the excitons.
This separation between the polaritonic modes in energy-momentum space is quantified by the Rabi splitting $\Delta_m = \min_{q} \Delta_m(q)$.
Differentiating $\Delta_m(q)$ with respect to $q$, we get the wavenumber that minimizes $\Delta_m (q)$, yielding
\begin{subequations}
    \begin{align}
        \Delta_{m}^{\text{TE}} &= 2 \sqrt{\alpha Z_{r} \hbar \omega_c p_{m} - \left(\alpha Z_{r}\dfrac{p_{m}}{E_{m}} \hbar\omega_{c} \right)^2},\label{eq:rabi_TE}
        \\
        \Delta_{m}^{\text{TM}} &= 2\sqrt{\alpha Z_{r}\dfrac{(\hbar\omega_c)^3 p_m}{E_m^2}\dfrac{1}{x_{m}} + \dfrac{E_{m}^{2}}{4} (1-x_{m})^{2}}, \label{eq:rabi_TM}
    \end{align}
where $x_{m}$ is the real solution of the cubic equation
\begin{equation}
    x_{m}^{3} - x_{m}^{2} - 2 \alpha Z_{r} \dfrac{(\hbar\omega_{c})^{3} p_m}{E_{m}^{4}} = 0.
\end{equation}
\end{subequations}

We verified that $x_m$ is close to $1$ across the entire bias range considered in this work.
With this observation, and noting that $\alpha^2 \ll \alpha$, it follows that the Rabi splittings can be approximated by
\begin{subequations}
    \begin{eqnarray}
        \Delta_m^{\text{TE}} \approx 2\sqrt{\alpha Z_r \hbar\omega_c p_m}, \label{eq:TEapprox}
        \\
        \Delta_m^{\text{TM}} \approx 2\sqrt{\alpha Z_r \frac{(\hbar \omega_c)^3 p_m}{E_m^2}}. \label{eq:TMapprox}
    \end{eqnarray}
\end{subequations}

\begin{figure*}[ht!]
    \includegraphics[width=1.35\columnwidth]{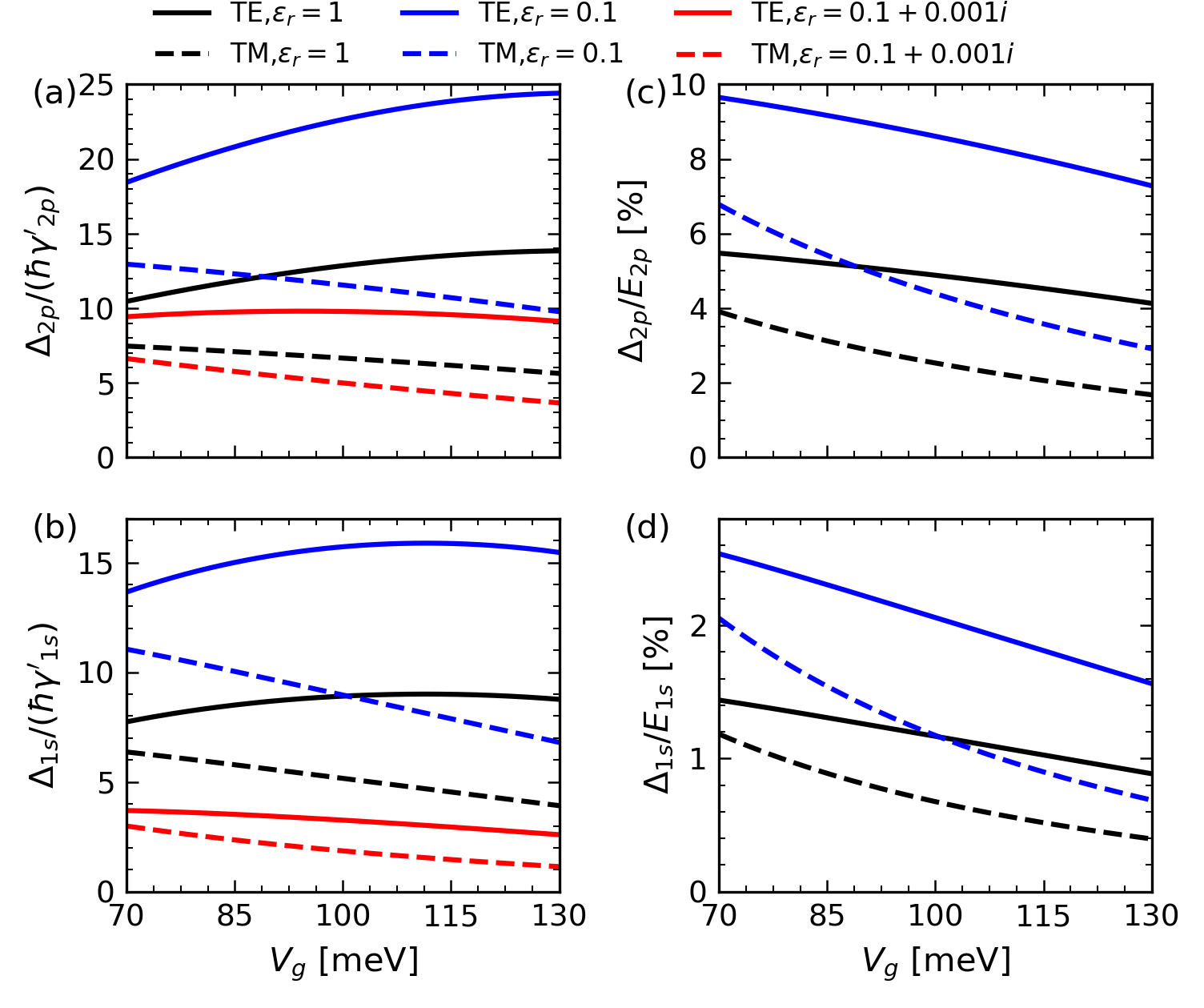}
    \caption{
    Rabi splitting $\Delta_{m}$ for each EP $1s$ (left) and $2p$ (right), as a function of the bias $V_g$.
        As indicated, the \ac{TE} modes \textcolor{black}{(solid lines)} always present higher splittings than the \ac{TM} modes \textcolor{black}{(dashed lines)}.
        The black curves represent the modes for $L=14$\,$\mu$m and a cavity filled with air ($\varepsilon_r=\mu_r=1$).
        The blue curves represent the modes for $L=45$\,$\mu$m and a cavity filled with an ENZ dielectric ($\epsilon_r=0.1$, $\mu_r=1$).
        \textcolor{black}{The red curves represent the modes for $L=45$ $\mu$m and a cavity filled with an ENZ dielectric with losses ($\epsilon_r=0.1+0.001i$, $\mu_r$=1).}
        Panels (a) and (b) show the Rabi splitting normalized by \textcolor{black}{$\hbar\gamma_{m}'$}, the polaritonic linewidths at the Rabi splitting point.
        Panels (c) and (d) show the Rabi splitting normalized by $E_{m}$, which increase linearly with the bias [see Fig.~\ref{fig2}].}
    \label{fig5}
\end{figure*}

Figs.~\ref{fig5}(a) and~\ref{fig5}(b) show the Rabi splitting for the \ac{TE} and \ac{TM} mode of each EP, as a function of bias voltage, for a cavity filled with air ($\epsilon_r=\mu_r=1$) and with an epsilon-near-zero (ENZ) dielectric ($\epsilon_r\sim 0$).
For means of direct comparison, we normalized the splittings by $\hbar \gamma^\prime_m = \hbar \gamma_m/2$, which is the effective linewidth $\hbar\gamma^{\prime \pm 1}_m(q)$ at the wavenumber value $q$ that corresponds to the Rabi splitting (see Fig.~\ref{fig4}).
We verified that the absolute value of the \ac{TE} Rabi splitting increases up to a certain peak, but since the exciton frequency also increases linearly and more steeply with the bias (see Fig.~\ref{fig1}), the normalized Rabi splitting shows a decreasing trend as shown in the lower panels of Fig.~\ref{fig5}.
The \ac{TE} modes present higher Rabi splittings in general, a behavior that can be understood directly from Eqs.~(\ref{eq:TEapprox}, \ref{eq:TMapprox}): the \ac{TM} Rabi splitting presents a dependency on $E_m^{-1}$.
Therefore, the \ac{TM} modes rapidly decrease with $V_g$, while the bias effect on the \ac{TE} modes is much softer and strictly increases the splitting.

To further increase the Rabi splitting of \ac{TE} modes, we reduced the dielectric constant to $\varepsilon_r=0.1$, representing an epsilon near-zero (ENZ) nondispersive lossless dielectric, which can be achievable through an engineered metamaterial~\cite{Sun2012}.
However, since reducing the dielectric constant increases the energies of the empty cavity photon modes [Eq.~\eqref{eq:empty cavity photon modes}], we had to increase the cavity width to $L=45\,\mu$m to maintain an anticrossing of the excitons energies and cavity modes. \textcolor{black}{We also present results for a lossy ENZ dielectric, with $\varepsilon_r=0.1+0.001i$. For this case, the increase of the Rabi splitting $\Delta_m$, for both TE and TM cases, is overshadowed by the increase of the EP linewidth $\gamma^j_m(q)$, calculated in this case by:
\begin{equation}
\gamma_m^j(q)=|X_m^j|^2\gamma_m+|Y_m|^2\gamma^\mathrm{ph}_q,
\end{equation}
with the cavity photon linewidth $\gamma^\mathrm{ph}_q$ obtained by the imaginary part of Eq. (\ref{eq:omegaqn}):
\begin{equation}
\gamma^\mathrm{ph}_q =-\mathrm{Im}\left[\frac{1}{\sqrt{\mu\epsilon}}\right] \sqrt{q^2 + \left( \frac{\pi}{L} \right)^2}.
\end{equation}
The results in Figs.~\ref{fig5}(a) and~\ref{fig5}(b) show that, in order to the addition of an ENZ to be truly effective, it should have negligible losses.}

\textcolor{black}{In} Figs.~\ref{fig5}(c) and~\ref{fig5}(d), we present the ratio $\Delta_m/E_m$ as a function of the bias, that can be directly compared with the inverse of the cavity quality factor $Q$, which will define the order of magnitude to be classified as a very large finesse.
If $\Delta_m/E_m$ is high enough, it would mean that a low quality factor suffices to satisfy the strong coupling condition $\Delta_m/E_m\gg 1/Q$ as long as we disregard the losses of the active material.
In this case, the $1s$ exciton at $\epsilon_r=1$ needs a minimum $Q$ of the order of $70$ ($1/Q \ll 0.014$) for the \ac{TE} mode and a higher Q ($\approx80$) for the \ac{TM} mode. When we increase the bias, we can see that the ratio $\Delta_m/E_m$  decreases,
reaching very low values in the \ac{TM} case while decreasing slower in the \ac{TE} case. This behavior can be understood directly from Eqs.~(\ref{eq:TEapprox}, \ref{eq:TMapprox}) and Fig.~\ref{fig2}(b): as we increase the bias, the exciton energy $E_m$ increases linearly,
and the Rabi splitting for the \ac{TM} mode decays with on $E_m^{-1}$. Therefore, the ratio $\Delta_m/E_m$ of \ac{TM} modes decay as $V_g^{-2}$, requiring very high $Q$ cavities even for biases as small as few hundreds of meV. 
The same does not occur for the \ac{TE} mode, whose dominant term in the square root depends only on $p_m$.
The same can be understood through the directions of the polariton \textcolor{black}{EM} field.
In the TE regime, the electric field is strictly in-plane, inducing stronger coupling with the exciton dipole due to its high radius despite the interlayer nature~\cite{park_nanolet10}.
In the TM regime, conversely, the electric field is out-of-plane, leading to weaker coupling with the excitons.

For the $2p$ exciton and $\epsilon_r=1$, the required cavity qualities for strong coupling are much lower.
At the bias range considered here, it starts at around $20$ ($1/Q \ll 0.05$) in the \ac{TE} case and $25$ ($1/Q \ll 0.04$) in the \ac{TM} case, which repeats the $1s$ behavior discussed just before.

At last, we briefly discuss the case of ENZ.
We can see that it requires cavities with lower $Q$ factors.
As the bias decreases, so will the exciton energies, bringing the Rabi splitting point closer to $(0,\omega_c)$, where the TE and TM modes degenerate into a single \textcolor{black}{EM} mode. Keeping $L$ constant, it follows that, at the limit $E_m = \hbar \omega_c$, Eqs.~(\ref{eq:TEapprox}, \ref{eq:TMapprox}) simplify to
\begin{eqnarray}
    \Delta^\mathrm{TE}_m \approx\Delta^\mathrm{TM}_m \approx 2\sqrt{\alpha Z_r \hbar \omega_c p_m },
\end{eqnarray}
\emph{i.e.}, the \ac{TE} and \ac{TM} modes are near degenerated.
Since $p_m$ slightly decreases with the bias, whereas $E_m$ linearly increases, a microcavity resonantly produced with a higher $\omega_c$ will have higher Rabi splittings at the cost of lower bias tunability because when $E_m<\hbar\omega_c$, we lose the resonant condition for the appearance of EP.

\section{Conclusion} \label{sec:conc}

The strong bias dependence of the excitonic optical properties of bilayer graphene provides an \emph{in situ} control of the anticrossing that emerges when the bilayer couples to the optical modes of a microcavity.
In this work, we focused on high $Q$ cavities, such that the main source of loss is the nonradiative channels of the exciton. The graphene bilayer has two main excitonic resonances, each one characterized by different requirements for the quality of the cavity for strong
coupling to be achieved. We showed that he resonance emerging from $2p$ exciton states couples more efficiently with light and is highly suitable for probing exciton--photon dynamics in cavities. 

Furthermore, we found strong coupling over the entire range of applied bias, that corresponds to exciton energies between 50--100\,meV, \emph{i.e.}, covering far-- and long--infrared frequencies. \ac{TE} modes have a higher $\Delta_m/\hbar\gamma'_m$ ratio than \ac{TM} ones, and consequently a higher $\Delta_m/E_m$ ratio, which indicates a more flexible requirement on the $Q$ quality factor of the cavity.

Apart from an empty (filled with air) cavity, we also considered the case of an ENZ material. Although it might be experimentally challenging to fill a cavity with an ENZ engineered metamaterial, it would be a potentially rewarding avenue that strongly increases the coupling of excitons and photons.

\section*{Acknowledgements}
The Center for Polariton-driven Light--Matter Interactions (POLIMA) is funded by the Danish National Research Foundation (Project No.~DNRF165).
N.~M.~R.~P. acknowledges support by the Portuguese Foundation for Science and Technology (FCT) in the framework of the Strategic Funding UIDB/04650/2020, COMPETE 2020, PORTUGAL 2020, FEDER, and through projects PTDC/FIS-MAC/2045/2021 and EXPL/FIS-MAC/0953/ 2021. N.~M.~R.~P. also acknowledges the Independent Research Fund Denmark (grant no. 2032-00045B) and the Danish National Research Foundation (Project No.~DNRF165)

V.~G.~M.~D. acknowledges a PhD. Scholarship from the Brazilian agency FAPESP (Fundação de Amparo à Pesquisa do Estado de São Paulo). A.~J.~C. were supported by CNPq (Conselho Nacional de Desenvolvimento Cient\'ifico e Tecnol\'ogico) Grant No. 423423/2021-5, 408144/2022-0, 315408/2021-9 and FAPESP under Grant No.~2022/08086-0 and a CAPES-PrInt scholarship.

\appendix
\section{Electromagnetic wave modes of exciton polaritons in a planar cavity}\label{app:method}

In this Appendix, we solve the Maxwell equations for the \textcolor{black}{EM} fields of a photon inside a planar cavity of dimension $L$, with two perfect metals at $z=\pm L/2$ as depicted in Fig.~\ref{fig1}.

Considering a single photon of frequency $\omega$, the Helmholtz equation
\begin{equation}
    \nabla^2 \mathbf{F} + \mu\epsilon\omega^2 \mathbf{F} = \mathbf{0}, \label{eq:hh}
\end{equation}
where $\mathbf{F}$ can be the electric field $\mathbf{E}$ or the magnetic field $\mathbf{B}$,
follows directly from the Maxwell equations in the absence of free charges.
From the translational invariance in the $xy$--plane, we have:
\begin{equation}
    \mathbf{F}(\mathbf{r},z) = \mathbf{f}(z) e^{i\mathbf{q}\cdot\mathbf{r}}, \label{eq:F-f}
\end{equation}
where $\mathbf{r}=x\mathbf{u}_x+y\mathbf{u}_y$ is the in-plane distance vector, and $\mathbf{q}=q_x\mathbf{u}_x+q_y\mathbf{u}_y$ is the in-plane wavevector.
The field amplitude $\mathbf f(z)$ must satisfy the reduced Helmholtz equation
\begin{equation}
    \partial_{z}^2 \mathbf{f}(z) + k^2 \mathbf{f}(z) = 0, \label{eq:red hh}
\end{equation}
where $k$ is the out-of-plane wavevector and $k^2 + q^2 = \mu\epsilon \omega^2$.
The time dependency $e^{i\omega t}$ has been factored out, such that all time derivatives must be recast as $\partial_{t}\rightarrow -i\omega$.
From the Gauss equation $\nabla\cdot\mathbf{F}=0$ with $\mathbf{f}(z)=f_{\perp}(z)\mathbf{u}_z+\mathbf{f}_{\parallel}(z)$ and $\mathbf{f}_{\parallel}(z)\cdot\mathbf{u}_z=0$, we have
\begin{equation}
    i\mathbf{q}\cdot\mathbf{f}_{\parallel}(z) + \partial_z f_{\perp}(z) = 0. \label{eq:f}
\end{equation}

From the Faraday equation $\nabla \times \mathbf{E} = i\omega \mathbf{B}$, we obtain
\begin{eqnarray}
    i\mathbf{q}\times \mathbf{e}_{\parallel}(z) + i e_{\perp}(z) \mathbf{q}\times\mathbf{u}_z + \mathbf{u}_z\times\partial_z \mathbf{e}_{\parallel}(z) =\nonumber\\
    i\omega \left[\mathbf{b}_{\parallel}(z) + b_{\perp}(z)\mathbf{u}_{z}\right].
\end{eqnarray}

Separating the in- and out-of-plane components, we obtain the equations%
\begin{subequations}
\begin{eqnarray}
    i\omega\mathbf{b}_{\parallel}(z) = ie_{\perp}(z)\mathbf{q}\times\mathbf{u}_{z} + \mathbf{u}_{z}\times \partial_{z} \mathbf{e}_{\parallel}(z),
    \\
    i\omega b_{\perp}(z)\mathbf{u}_{z} = i\mathbf{q}\times \mathbf{e}_{\parallel}(z). \label{eq:bz}
\end{eqnarray}
\end{subequations}

Applying the cross product $\mathbf{q}\times$ on both sides of Eq.~\eqref{eq:bz}, we get
\begin{equation}
    i\omega b_{\perp}(z) \mathbf{q}\times\mathbf{u}_{z} = i\left[\mathbf{q}\cdot \mathbf{e}_{\parallel}(z)\right] \mathbf{q} - iq^2 \mathbf{e}_{\parallel}(z),
\end{equation}
now using Eq.~\eqref{eq:f}:
\begin{equation}
    i\omega b_{\perp}(z) \mathbf{q}\times\mathbf{u}_z = -\partial_{z} e_{\perp}(z)\mathbf{q}-iq^2 \mathbf{e}_{\parallel}(z),
\end{equation}
and we finally obtain an expression for the in-plane component of the electric field amplitude:
\begin{subequations}    
\begin{equation}
    \mathbf{e}_{\parallel}(z) = \dfrac{-\partial_{z}e_{\perp}(z)\mathbf{q}-i\omega(\mathbf{q}\times\mathbf{u}_{z}) b_{\perp}(z)}{iq^2}. \label{eq:e par}
\end{equation}

Through an entirely analogous procedure, we obtain the in-plane component of the magnetic field amplitude, starting from the Maxwell-Amp\`{e}re equation $\nabla \times \mathbf{B} = -i\omega\mu\epsilon\mathbf{E}$:
\begin{equation}
    \mathbf{b}_{\parallel}(z) = \dfrac{-\partial_{z}b_{\perp}(z)\mathbf{q}+i\omega\mu\epsilon(\mathbf{q}\times\mathbf{u}_{z}) e_{\perp}(z)}{iq^2}. \label{eq:b par}
\end{equation}
\end{subequations}

From the boundary conditions at the metallic interfaces, Eq.~\eqref{eq:e par} gives
\begin{equation}
    \dfrac{-\partial_z e_\perp (\pm L/2) \mathbf q - i\omega (\mathbf q \times \mathbf u_z) b_\perp (\pm L/2)}{iq^2} = \mathbf{0}.
\end{equation}
Therefore, for the in- and out-of-plane components we have
\begin{eqnarray}
    \partial_{z} e_\perp (\pm L/2) = 0,
    \\
    b_\perp (\pm L/2) = 0.
\end{eqnarray}
These results, together with the reduced Helmholtz equation~\eqref{eq:red hh}, allow us to write $e_{\perp}^{\pm}(z) = \pm E_0 \cos\theta_{n}^{\pm}$ and $b_{\perp}^{\pm}(z) = \pm B_0 \sin\theta_{n}^{\pm}$ ($\theta_{n}^{\pm}= k(z\mp L/2)$), from which we can also explicitly derive the in-plane amplitudes using Eqs.~(\ref{eq:e par}, \ref{eq:b par}):
\begin{subequations}\label{eq:eig eb}
\begin{eqnarray}
    \dfrac{\mathbf{e}_{\parallel}^{\pm}(z)}{\sin\theta_{n}^{\pm}} = \pm \left(i k \dfrac{\mathbf{q}}{q^2} E_0 - \omega \dfrac{\mathbf{q}\times\mathbf{u}_z}{q^2} B_0\right), \label{eq:eig e}
    \\
    \dfrac{\mathbf{b}_{\parallel}^{\pm}(z)}{\cos\theta_{n}^{\pm}} = \mp \left(i k \dfrac{\mathbf{q}}{q^2} B_0 - \mu\epsilon\omega \dfrac{\mathbf{q}\times\mathbf{u}_{z}}{q^2} E_0\right). \label{eq:eig b}
\end{eqnarray}
\end{subequations}

Now applying the boundary condition at the \ac{2D} material interface ($z=0$)
\begin{equation}
    \mathbf{u}_{z} \times (\mathbf{b}^{+}-\mathbf{b}^{-}) = \mu\sigma(\omega) \mathbf{e}_{\parallel}^{\pm},
\end{equation}
it follows that
\begin{eqnarray}
    -2 \left(i k \dfrac{\mathbf{u}_{z}\times \mathbf{q}}{q^{2}} B_0 -\mu\epsilon\omega\dfrac{\mathbf{q}}{q^{2}}E_0 \right) \cos(kL/2)
    = \nonumber\\
    -\mu\sigma(\omega) \left(ik \dfrac{\mathbf{q}}{q^2} E_0 - \omega \dfrac{\mathbf{q}\times\mathbf{u}_z}{q^2}B_0\right)\sin(kL/2).
\end{eqnarray}
Separating the orthogonal components, we end up with
\begin{subequations}
\begin{eqnarray}
    \textcolor{black}{-}2ik \cos(kL/2) = \mu\omega\sigma(\omega)\sin(kL/2),
    \\
    2\epsilon\omega\cos(kL/2) = \sigma(\omega)ik\sin(kL/2).
\end{eqnarray}
\end{subequations}
\textcolor{black}{
To rearrange these equations as presented in the main text, we use the fine structure constant $\alpha = \sigma_0/(\pi c \epsilon_0)$:
\begin{subequations}
\begin{eqnarray}
    \frac{kc}{\omega}\cot(kL/2) = -\frac{i\mu_r \pi\alpha}{2}\frac{\sigma(\omega)}{\sigma_0},
    \\
    \frac{\omega}{kc}\cot(kL/2) = \frac{i\pi\alpha}{2\epsilon_r}\frac{\sigma(\omega)}{\sigma_0}.
\end{eqnarray}
\end{subequations}
}

If there was no \ac{2D} material at $z=0$, the system could be regarded as one single dielectric of length $L$, 
and the fields labeled with $\pm$ would be identical.
This leads to
\begin{equation}
    k = \dfrac{n\pi}{L},
\end{equation}
where $n$ is an integer and the dispersion relation of modes in the empty cavity is given by
\begin{equation}
    \omega_{qn} = \dfrac{1}{\sqrt{\mu\epsilon}} \sqrt{q^2 + \left( \dfrac{n\pi}{L} \right)^2}, \label{eq:empty cavity photon modes}
\end{equation}
as presented in the main text.

\section{Quantization of the photon electromagnetic field in a planar cavity}\label{app.quantization}

In this Appendix, we initially consider an empty cavity, with only the boundary conditions given by the perfect mirrors.
To derive the Hamiltonian of cavity photons, we will use the vector potential $\mathbf A(\mathbf r,t)$ in the Weyl gauge, $\mathbf{E}(\mathbf r,t) = -\partial_{t}\mathbf{A}(\mathbf r,t)$.
\textcolor{black}{
As shown in App.~\ref{app:method}, the EM fields have translational invariance in the in-plane direction and are quantized in the $z$--direction, with a wavevector $k_z=n\pi/L$.
Therefore, we can write the electric and magnetic fields as $\mathbf{E}(\mathbf{r},t) = \mathbf{e}(z) e^{i(\mathbf{q}\cdot\mathbf{r}-\omega t)}$ and $\mathbf{B}(\mathbf{r},t) = \mathbf{b}(z) e^{i(\mathbf{q}\cdot\mathbf{r}-\omega t)}$, and derive expressions for $\mathbf{e}(z)$ and $\mathbf{b}(z)$ [Eqs.~\eqref{eq:eig eb}].
Using this result, we will derive equations for the EM eigenmodes $\mathbf{u}_j(z)$ ($j=\text{TE}, \text{TM}$), taking into account that the TE mode occurs when $E_0 = 0$ [Eq.~\eqref{eq:TE}], and the \ac{TM} mode, when $B_0 = 0$ [Eq.~\eqref{eq:TM}].
}
At last, we will use the resultant eigenmodes to quantize the \textcolor{black}{EM} fields of the cavity photon, and arrive at an expression for the Hamiltonian of cavity photons.

Considering an empty cavity, \textcolor{black}{Eqs.~\eqref{eq:eig eb}} become
\begin{subequations}\label{eq:eig eb 0}
\begin{eqnarray}
    \mathbf e(z) = E_0 \cos\theta_n \mathbf u_z + \nonumber\\ 
    \left(i k \dfrac{\mathbf{q}}{q^2} E_0 - \omega \dfrac{\mathbf{q}\times\mathbf{u}_z}{q^2} B_0\right) \sin\theta_n, \label{eq:eig e 0}
    \\
    \mathbf b(z) = B_0 \sin\theta_n \mathbf u_z + \nonumber\\
    \left(i k \dfrac{\mathbf{q}}{q^2} B_0 - \mu\epsilon\omega \dfrac{\mathbf{q}\times\mathbf{u}_{z}}{q^2} E_0\right) \cos\theta_n, \label{eq:eig b 0}
\end{eqnarray}
\end{subequations}
where $\theta_n = (n\pi/L)(z+L/2)$.
In the Weyl gauge, we get $\mathbf e(z) = i\omega \mathbf a(z)$, and we can define the eigenmodes
\begin{subequations}\label{eq:u}
\begin{align}
    \mathbf u_{\text{TE}}(z) &= \sqrt{\frac{2}{L}}
    i\dfrac{\mathbf{q}\times\mathbf{u}_z}{q} \sin\theta_{n}, \label{eq:uTE}
    \\
    \mathbf u_{\text{TM}}(z) &= \sqrt{\frac{2}{\mu\epsilon L}} \left( \dfrac{n\pi}{L} \dfrac{\mathbf{q}}{\omega q} \sin\theta_{n} - i\dfrac{q}{\omega} \cos\theta_{n}\mathbf{u}_{z} \right). \label{eq:uTM}
\end{align}
\end{subequations}
The vector potential is then written as a decomposition on these eigenmodes:
\begin{equation}
    \mathbf{A}(\mathbf{r},z) = \sum_{\mathbf{q} n j} \sqrt{\dfrac{\hbar}{2\epsilon S \omega_{qn}}} b_{\mathbf{q}nj} \mathbf{u}_{\mathbf{q}nj}(z) e^{i\mathbf{q}\cdot\mathbf{r}} + \text{H.c.}, \label{eq:Aquant}
\end{equation}
where $S$ is the total surface area of the \ac{2D} material, $\omega_{qn}$ is the empty cavity photon frequency~\eqref{eq:empty cavity photon modes}, and $b_{\mathbf q n}$ is the annihilation operator of the cavity photon, which follows the bosonic commutation rules.
Using this quantized vector potential, the Hamiltonian follows from $\mathbf{E}=-\partial_{t}\mathbf{A}$ and $\mathbf{B} = \nabla \times \mathbf{A}$:
\begin{equation}
    H_{cav} = \int \text{d}r^2 \text{d}z \left[\dfrac{\epsilon}{2} (\partial_{t} \mathbf{A})^2 + \dfrac{1}{2\mu} (\nabla \times \mathbf{A})^2\right].
\end{equation}
Using Eq.~\eqref{eq:Aquant}, one can show, after several algebraic steps, that
\begin{equation}
    H_{cav} = \sum_{\mathbf{q}nj} \hbar\omega_{qn}\left(b^{\dagger}_{\mathbf{q}nj}b_{\mathbf{q}nj} + \dfrac{1}{2}\right).
\end{equation}

\section{Electron--photon interaction in a planar cavity}\label{app.electron-cavity photon}

In this Appendix, we will derive a Hamiltonian for the electron--photon interaction inside the planar cavity assuming dipolar coupling.
Then we project this Hamiltonian on the basis of excitons and photons and extract an expression for the exciton--photon coupling function.

In order to describe the electron--photon interaction, We start with the Wannier Hamiltonian:
\begin{equation}
    H_{wan} = \sum_{\mathbf{k}\lambda} E_{\mathbf k \lambda} a^{\dagger}_{\mathbf{k}\lambda} a_{\mathbf{k}\lambda} + H_{ee},
\end{equation}
where $a_{\mathbf{k}\lambda}$ is the fermionic operator of an electron in the band $\lambda$ with momentum $\mathbf{k}$ and energy $E_{\mathbf k \lambda}$, and $H_{ee}$ is the electron-electron interaction.

In the dipole approximation, we assume the photons and electrons interact via coupling of the photon electric field with the electric dipole moment associated to an excited electron and a hole.
The corresponding Hamiltonian is
\begin{equation}
    H_{dip} = -e \mathbf{E}\cdot\mathbf{r}, \label{eq:Hdip0}
\end{equation}
where $\mathbf{E}$ and $\mathbf{r}$ are the electric field and position operators applied to a point on a \ac{2D} material placed at $z=z_{0}$.
Using $\mathbf{E} = i\omega\mathbf A$ and Eq.~\eqref{eq:Aquant}, we get
\begin{flalign}
    H_{dip} = -e \sum_{\mathbf{q}nj} i\sqrt{\dfrac{\hbar\omega_{qn}}{2\epsilon S}} b_{\mathbf{q}nj} \mathbf{u}_{\mathbf{q}nj}(z_0) e^{i\mathbf{q}\cdot\mathbf{r}} \cdot\mathbf{r} +
    \text{H.c.}. \label{eq:Hdip}
\end{flalign}

Neglecting the terms from the continuum, the Wannier Hamiltonian becomes diagonal in the excitonic basis:
\begin{equation}
    H_{wan} | \mathbf{q}m \rangle = E_{\mathbf q m} | \mathbf{q}m \rangle, \label{eq:H0 wannier}
\end{equation}
where $E_{\mathbf q m}$ is the exciton energy, and
\begin{equation}
    | \mathbf{q}m \rangle = \sum_{\mathbf{k}} \phi_{\mathbf{q}m}(\mathbf{k}) a_{\mathbf{k}+\mathbf{q},c}^{\dagger} a_{\mathbf{k}v} | \psi_0 \rangle, \label{eq:exciton state}
\end{equation}
where $\phi_{\mathbf{q}m}(\mathbf{k})$ is the exciton wavefunction with center-of-mass momentum $\mathbf{q}$ and quantum numbers $m$, and $|\psi_0\rangle$ is ground state of system, \textit{i.e.}, the excitonic vacuum.
Now recasting Eq.~\eqref{eq:H0 wannier} in the second quantization picture, we have
\begin{equation}
    H_{wan} = \sum_{\mathbf{q}m} E_{\mathbf q m} c_{\mathbf{q}m}^{\dagger} c_{\mathbf{q}m},
\end{equation}
where the exciton operator is
\begin{equation}
    c^\dagger_{\mathbf{q}m} = \sum_{\mathbf{k}} \phi_{\mathbf{q}m}(\mathbf{k}) a_{\mathbf{k}+\mathbf{q},c}^{\dagger} a_{\mathbf{k}v} ,\label{eq:exc.}
\end{equation}
and $S$ is the surface area of the \ac{2D} material.

The next step is to rewrite the dipole Hamiltonian in the exciton--photon basis:
\begin{equation}
    H_{dip} = \sum_{\mathbf{q} m n j} g_{mnj}(\mathbf{q}) c_{\mathbf{q}m}^{\dagger} b_{\mathbf{q}nj} + \text{H.c.},
\end{equation}
where we neglect the intraband transitions and the continuum part of the energy spectrum.
The exciton--photon coupling is then given by matrix element
\begin{equation}
    g_{mnj}(\mathbf{q}) = \langle \mathbf{0}_{\text{ph}};\mathbf{q},m | H_{dip} | \mathbf{q},n,j;\mathbf{0}_{\text{ex}} \rangle,
\end{equation}
which corresponds to the annihilation of a photon of in-plane wavenumber $\mathbf{q}$, cavity quantum number $n$ and polarization $j$, followed by the creation of an exciton of center-of-mass momentum $\mathbf{q}$ and quantum numbers $m$.
Substituting the dipole Hamiltonian~\eqref{eq:Hdip}, we get
\begin{eqnarray}
    g_{mnj}(\mathbf{q}) = -e \sum_{\mathbf q' n' j'} i\sqrt{\dfrac{\hbar\omega_{q'n'}}{2\epsilon S}} \mathbf{u}_{\mathbf{q}'n'j'}(z_0) \cdot \nonumber\\
    \langle \mathbf{0}_{\text{ph}};\mathbf{q},m | b_{\mathbf{q}'n'j'} e^{i\mathbf{q}'\cdot\mathbf{r}} \mathbf{r} | \mathbf{q},n,j;\mathbf{0}_{\text{ex}} \rangle.
\end{eqnarray}
Now expanding the exciton state~\eqref{eq:exciton state} and using $b_{\mathbf{q}'n'j'} | \mathbf{q},n,j;\mathbf{0}_{\text{ex}} \rangle = \delta_{\mathbf q \mathbf q'} \delta_{nn'} \delta_{jj'}|\psi_0\rangle$:
\begin{eqnarray}
    g_{mnj}(\mathbf{q}) = -i e \sqrt{\dfrac{\hbar\omega_{qn}}{2\epsilon S}} \sum_{\mathbf{k}} \phi^{*}_{\mathbf{q}m}(\mathbf{k}) \times \nonumber\\
    \mathbf{u}_{\mathbf{q}nj}(z_0)\cdot \langle \psi_0 | a^{\dagger}_{\mathbf{k}v} a_{\mathbf{k}+\mathbf{q},c} e^{i\mathbf{q}\cdot\mathbf{r}} \mathbf{r} |\psi_0\rangle. \label{eq:g1}
\end{eqnarray}

The matrix element in Eq.~\eqref{eq:g1} is closely related to the excitonic transition dipole moment.
To show this, we first write the electron operator in the position basis:
\begin{equation}
    a_{\mathbf{k}\lambda} = \dfrac{1}{\sqrt{N}} \sum_{\mathbf{R}\boldsymbol{\delta}} e^{-i\mathbf{k}\cdot(\mathbf{R}+\boldsymbol{\delta})} v^{*}_{\mathbf{k}\lambda\boldsymbol\delta} a_{\mathbf{R}+\boldsymbol{\delta}},
\end{equation}
where $N\gg 1$ is the number of unit cells, $\mathbf{R}$ and $\boldsymbol{\delta}$ are the lattice and basis vectors in configuration space, respectively, and $v_{\mathbf{k}\lambda\boldsymbol{\delta}}$ are the Bloch functions.
In the long wavelength limit, the photon wavelength is much greater than the unit cell dimensions, therefore we can use $v_{\mathbf{k}+\mathbf{q},\lambda\boldsymbol\delta}\approx v_{\mathbf{k}\lambda \boldsymbol{\delta}}$.
Additionally, since $e^{i\mathbf{q}\cdot\mathbf{r}}|\mathbf{R}+\boldsymbol{\delta}\rangle = e^{i\mathbf{q}\cdot(\mathbf{R}+\boldsymbol{\delta})}|\mathbf{R}+\boldsymbol{\delta}\rangle $, we have $a_{\mathbf{k}+\mathbf {q},\lambda}e^{i\mathbf{q}\cdot\mathbf{r}}\approx a_{\mathbf{k}\lambda}$.
Thus,
\begin{align}
    -e\langle \psi_0 | a^{\dagger}_{\mathbf{k}v} a_{\mathbf{k}+\mathbf{q},c} \mathbf{r}e^{i\mathbf{q}\cdot\mathbf{r}}|\psi_0\rangle &\approx -e\langle\psi_0| a^{\dagger}_{\mathbf{k}v} a_{\mathbf{k}c} \mathbf{r} |\psi_0\rangle \nonumber\\
    &= e\langle \mathbf{k} c | \mathbf{r} | \mathbf{k} v \rangle, \label{eq:bra-ket term}
\end{align}
which we recognize as the excitonic transition dipole moment $\mathbf d_{cv}^{\mathbf k \mathbf k}$.
Using Eq.~\eqref{eq:bra-ket term} and introducing the excitonic transition dipole moment
\begin{equation}
    \mathbf{d}_{\mathbf{q}m} = \dfrac{1}{\sqrt{S}}\sum_{\mathbf{k}}\phi_{\mathbf{q}m}^{*}(\mathbf{k}) \mathbf{d}_{cv}^{\mathbf{k}+\mathbf{q},\mathbf k},
\end{equation}
the coupling function becomes
\begin{align}
    g_{mnj}(\mathbf{q}) &= i \sqrt{\dfrac{\hbar\omega_{qn}}{2\epsilon}} \mathbf{u}_{\mathbf{q}nj}(z_0) \cdot \mathbf{d}_{\mathbf{0}m} \label{eq:g2}.
\end{align}

\section{Elliott formula from exciton--photon coupling} \label{app:elliot}

In this Appendix, we derive the Elliott formula for the optical conductivity [Eq.~\eqref{eq:elliot}].
We start by considering the classical polarization vector from the expectation value of the dipole operator, then write the Heisenberg equation of motion considering a Hamiltonian of excitons and a dipolar interaction of the excitons with the classical electric field $\mathbf E$ of the photons.
Using this equation of motion, we obtain the expectation value of the exciton operator, which can be used to derive an expression for the polarization vector that can be related to the optical conductivity through the polarization current $\mathbf J_p = -i\omega\mathbf P = \boldsymbol \sigma\cdot\mathbf E$.

We can calculate the total polarization vector due to excitons, starting with the expectation value of the electric dipole moment operator for the excitonic contributions
\begin{equation}
\mathbf{P} =e\langle \mathbf{r}\rangle.
\end{equation}
Writing the position operator $\mathbf r$ in the electron--hole basis, considering the Heisenberg picture, we have
\begin{equation}
\mathbf{P} = \sum_\mathbf{k} \mathbf{d}_{cv}^{\mathbf{k}\mathbf k} \langle \hat{a}_{\mathbf{k}c}^\dagger \hat{a}_{\mathbf{k}v} \rangle + \text{H.c.} \label{eq:P1}
\end{equation}
Neglecting the terms from the continuum, we can invert the relation (\ref{eq:exc.}):
\begin{equation}
a^\dagger_{\mathbf{k}+\mathbf{q},c}a_{\mathbf{k}v} = \sum_{m} \phi^*_{\mathbf{q}m}(\mathbf{k})c^\dagger_{\mathbf{q}m},
\end{equation}
and use it to rewrite the polarization in the excitonic basis:
\begin{align}
\mathbf{P} &= \sum_m \sum_{\mathbf{k}} \phi^*_{\mathbf{0}m}(\mathbf{k})\mathbf{d}^{\mathbf{k}\mathbf{k}}_{cv} \langle \hat{c}^\dagger_{\mathbf{0}m} \rangle + \text{H.c.}\nonumber\\
&= \sum_{m} \mathbf{d}_{\mathbf{0}m} \langle c^{\dagger}_{\mathbf{0}m} \rangle + \text{H.c.} \label{eq:P2}
\end{align}
Therefore, we can redefine the excitonic electric dipole moment operator as
\begin{equation}
    \boldsymbol{\mathcal{P}} = e\mathbf r = \sum_{m} \mathbf{d}_{\mathbf 0 m} c^\dagger_{\mathbf 0 m} + \text{H.c.} \label{eq:P}
\end{equation}

Now we introduce the dipolar coupling with a classical electric field $\mathbf{E}$:
\begin{equation}
H'_{dip}=-\mathbf{E}\cdot\boldsymbol{\mathcal{P}}= -\mathbf{E}\cdot \sum_m \left(\mathbf d_{\mathbf 0 m} c^\dagger_{\mathbf 0 m} + \mathbf d^*_{\mathbf 0 m} c_{\mathbf 0 m}\right).
\end{equation}
In the excitonic basis, the total Hamiltonian---now considering a classical field for the photons---becomes
\begin{align}
H' &= H_{wan} + H'_{dip} \nonumber\\
&= \sum_{\mathbf{q}m} E_m(\mathbf{q}) \hat{c}^\dagger_{\mathbf{q}m}\hat{c}_{\mathbf{q}m}-\nonumber\\
& \mathbf{E}\cdot \sum_m \left( \mathbf{d}_{\mathbf{0}m} c^\dagger_{\mathbf{0}m}+\mathbf{d}^*_{\mathbf{0}m}c_{\mathbf{0}m}\right).
\end{align}
The Heisenberg equation of motion gives us
\begin{equation}
i\hbar \frac{d}{dt}\langle\hat{c}^\dagger_{\mathbf{0}m}\rangle =\langle[H',\hat{c}^\dagger_{\mathbf{0}m} ]\rangle=E_m\langle\hat{c}^\dagger_{\mathbf{0}m}\rangle -\mathbf{E}\cdot \mathbf{d}^*_{\mathbf{0}m}.
\end{equation}
Therefore, we obtain the solution in linear order
\begin{equation}
\langle c^\dagger_{\mathbf{0}m} \rangle = -\frac{\mathbf{E}\cdot \mathbf{d}^*_{\mathbf{0}m}}{\hbar\omega - E_m}, \label{eq:c0m}
\end{equation}
which is what we need to substitute on Eq.~\eqref{eq:P2} to carry on. 
The additional Hermitian conjugate term on Eq.~\eqref{eq:P2} depends on $\langle c_{\mathbf 0 m} \rangle$ and yields a factor $(\hbar\omega + E_m)^{-1}$, \textit{i.e.}, a rapidly oscillating term that can be safely neglected in light of the rotating wave approximation.
Substituting Eq.~\eqref{eq:c0m} back into Eq.~\eqref{eq:P2}, and using the polarization current definition $\mathbf{J}_p=\partial_t \mathbf{P}=-i\omega\mathbf P$, we obtain%
\begin{align}
    \mathbf{J}_p &= i\omega \sum_{m} \mathbf{d}_{\mathbf{0}m}
    \dfrac{\mathbf{E}\cdot\mathbf{d}^{*}_{\mathbf{0}m}}{\hbar\omega-E_m}
    \nonumber\\
    &= i\omega \sum_{m} \dfrac{\mathbf{d}_{\mathbf{0}m}\otimes \mathbf{d}^{*}_{\mathbf{0}m}}{\hbar\omega-E_m} \cdot\mathbf{E},
\end{align}
where we used the symbol $\otimes$ to denote an outer product.

Now using the relation between current density and optical conductivity $\mathbf{J}_p=\boldsymbol\sigma(\omega)\mathbf{E}$, we get
\begin{equation}
    \boldsymbol\sigma(\omega) = i\omega \sum_{m} \dfrac{\mathbf d_{\mathbf 0 m} \otimes \mathbf d_{\mathbf 0 m}^*}{\hbar\omega-E_m},
\end{equation}
the so-called Elliott formula~\cite{Chaves_2017,ElliotOpticalAbsorptionExcitons}.
To rewrite it as presented in the main text, we need to perform some simplifications.
First, we verified that, for bilayer graphene, we have $\sigma_{xx} \approx \sigma_{yy}$ and $\sigma_{xy} \approx \sigma_{yx} \approx 0$, which is a reasonable finding since bilayer graphene manifests weak anisotropy in momentum space.
Moreover, we can safely neglect the $z$--component of the dipole since the excitons of bilayer graphene have radii much greater than its interlayer distance.
In addition, since we are in the dipolar approximation, the only optically active component of the exciton dipole is the one parallel to $\mathbf E$.
These considerations allow us to can recast the optical conductivity tensor of bilayer graphene to a scalar:
\begin{equation}
    \sigma(\omega) = i\omega \sum_{m} \frac{d^2}{\hbar\omega - E_m},
\end{equation}
where $d$ is the magnitude of the dipole projected in the electric field direction, neglecting any $z$--component contribution.
To rewrite this as presented in Eq.~\eqref{eq:elliot}, we perform the substitution $E_m \rightarrow E_m - i\hbar\gamma_m$ to account for the radiative decay of the exciton and define the oscillator strength
\begin{equation}
    p_m = \frac{E_m}{\hbar\sigma_0} d^2, \label{eq:pm}
\end{equation}
that links the classical and quantum formalisms.

Examining now Eqs.~\eqref{eq:u} at $z=z_0=0$ and neglecting the $z$--component of the dipole, we can write
\begin{equation}
    \mathbf u_{\mathbf q nj}(0) \cdot \mathbf d_{\mathbf 0 m} = u_j  d \sin\left(\frac{n\pi}{2}\right), \label{eq:umod}
\end{equation}
where $u_{\text{TE}} = i\sqrt{2/L}$ and $u_{\text{TM}} = n (\omega_{c}/\omega_{qn}) \sqrt{2/L}$.
Substituting Eq.~\eqref{eq:umod} in Eq.~\eqref{eq:g2}, we get
\begin{equation}\label{eq:g3}
    g_{mnj}(\mathbf q) = i \sqrt{\frac{\hbar \omega_{qn}}{2\epsilon}} u_j d\sin\left(\frac{n\pi}{2}\right).
\end{equation}
Using the oscillator strength~\eqref{eq:pm}, we arrive at
\begin{equation}
    g_{mnj}(\mathbf q) = i \sqrt{\frac{\hbar \omega_{qn}}{2\epsilon}} u_j \sqrt{\hbar\sigma_0 \frac{p_m}{E_m}} \sin\left(\frac{n\pi}{2}\right).
\end{equation}
Recognizing $\sigma_0 = \pi\alpha c \epsilon_0$, we have
\begin{align}
    g_{mnj}(\mathbf q) &= i \sqrt{\frac{\hbar \omega_{qn}}{2\epsilon_r} \hbar \pi \alpha c \frac{p_m}{E_m}} u_j \sin\left(\frac{n\pi}{2}\right) \nonumber\\
    &= i \sqrt{\frac{\hbar \omega_{qn}}{\epsilon_r} \hbar \alpha \frac{\pi c}{L} \frac{p_m}{E_m}} \frac{u_j}{\sqrt{2/L}} \sin\left(\frac{n\pi}{2}\right) \nonumber\\
    &= i \sqrt{\hbar \omega_{qn} Z_r \hbar \alpha \omega_c \frac{p_m}{E_m}} \frac{u_j}{\sqrt{2/L}} \sin\left(\frac{n\pi}{2}\right).
\end{align}
Rearranging,
\begin{equation}
    g_{mnj}(\mathbf q) = i \hbar \bar{u}_j \sqrt{\omega_{qn}\omega_c} \sqrt{\alpha Z_r \frac{p_m}{E_m}} \sin\left(\frac{n\pi}{2}\right),
\end{equation}
where $\bar{u}_j = u_j \sqrt{L/2}$ is a dimensionless factor. Writing $u_0^j$ explicitly for each $j=\text{TE},\text{TM}$ and particularizing the results for $n=1$, we obtain Eqs.~(\ref{eq:gTE}, \ref{eq:gTM}).

\section{\textcolor{black}{Numerical minimization of $|f^{\text{TE}}(q,\omega)|$ and $|f^{\text{TM}}(q,\omega)|$}}\label{app:numerical minimization}

\textcolor{black}{
The numerical solution of Eqs.(1) is performed in two steps, which we describe in this Appendix.
First, we define the functions
\begin{subequations}
\begin{eqnarray}
    f^{\text{TE}}(q,\omega) = \frac{kc}{\omega} \cot(kL/2) - i\frac{\mu_r \pi \alpha}{2} \frac{\sigma(\omega)}{\sigma_0},
    \\
    f^{\text{TM}}(q,\omega) = \frac{\omega}{kc} \cot(kL/2) - i\frac{\pi\alpha}{2\epsilon_r} \frac{\sigma(\omega)}{\sigma_0},
\end{eqnarray}
\end{subequations}
as presented in main text.
Clearly, Eqs.~(1) correspond to the roots of $f^j(q,\omega)$.
The color maps of $\log|f^j(q,\omega)|$ for real $q$ and $\omega $ are presented in Fig.~3(a) and (b), showing a unique minimum at three frequency ranges $\omega < \omega_{1s}$,  $\omega_{1s} < \omega < \omega_{2p}$, and $\omega > \omega_{2p}$, where each polaritonic branch is contained.
Therefore, we performed standard numerical minimization techniques to calculate the real frequency $\omega(q)$ that minimizes $g^j(q) = \log |f^j(q,\omega(q))|$, as a function of $q$, for each frequency range.
We call the solutions of this first minimization step $\omega_i^{j,(0)}(q)$ for each polaritonic branch $i=1,2,3$.
}

\textcolor{black}{
We begin the second step by defining $h^j(q,\gamma)~=~\log |f^j(q,\omega_i(q)~-~i\gamma)|$.
As discussed in the main text, the solutions of $f^j(q,\omega-i\gamma)=0$ must only deviate from $\omega_i(q)$ close to the exciton frequencies $\omega_{n}$.
We examined $h^j(q,\gamma)$ in the space of $\gamma$ vs. $q$ and verified that its roots stay close to $\gamma=0$ for $q$ values where $\omega_i(q)$ is far from the exciton frequencies, and it shifts to monotonically increasing $\gamma$ when $\omega_i(q)$ approaches $\omega_n$.
Hence, we performed standard numerical minimization techniques of multivariate functions to calculate, for each of the three polaritonic branches, the pair  $\{\omega(q),\gamma(q)\}$ that minimizes $\log|f^j(q,\omega(q)-i\gamma(q))|$, as a function of $q$, using the initial guesses $(\omega_i(q),0)$ obtained in the previous step.
We call the solutions of this second minimization step $\{\omega^{j,(1)}_{i}(q), \gamma^{j,(1)}_{i}(q)\}$.
}

\end{document}